# The effect of aging in thin films in the picosecond sonar experiment

*Petra Veselá\*, Martina Hlubučková, Vít Kanclíř, Jan Václavík and Karel Žídek*

*Institute of Plasma Physics of the Czech Academy of Sciences, U Slovanky 2525/1a, Prague, 182 00 Czech Republic*
\*vesela@ipp.cas.cz

## Abstract

This work examines aging effects in thin-film stacks studied by picosecond acoustics experiments. The method uses a strain wave created by the absorption of a laser pulse in the top metal transducer layer to study the inner structure and properties of multilayers. We show that significant changes in measured signals developed over time, rapidly after sample deposition and continuing for months. Most apparent was the evolution of the thermal exponential background of the signal, but close inspection showed modifications also in the shape of the echoes and the phase of the Brillouin oscillations. We revealed that these changes originate from the titanium transducer layer exposed to the ambient atmosphere. The aging process involves both - irreversible Ti oxidation and reversible gas/water vapor adsorption. Importantly, these aging effects can be eliminated by passivating the Ti layer with a thin 10 nm $Si_3N_4$ coating, providing a simple solution for ensuring long-term measurement stability.

# 1. Introduction

Optical interference thin films are important in many applications and research fields, which has triggered the need for their precise characterization [1]. Picosecond acoustics (also called ps sonar) is a well-suited method for the characterization of these films, as it can provide information about film thickness, elastic and optical properties of thin-film materials, or information about film interfaces, such as layer adhesion [2, 3]. This experiment also finds its place in basic research, e.g., for studying electron-phonon and electron-electron scattering [4], for studying nanostructures [5], or for imaging of biological samples [6]. As a non-destructive method, ps acoustics is well suited for studying changes in materials over time and for detecting hidden flaws or structural issues in materials [7].

In this technique, a short laser pulse (pump pulse) triggers an acoustic wave propagating through a material. Another delayed pulse (probe pulse) can then be employed to track the acoustic wave by monitoring changes in its reflectivity. An important component of the experiment is an opto-acoustic transducer. Typically, it is a nm-thin metal film deposited as a top layer, which absorbs the pump pulse and triggers the acoustic wave. Moreover, the thin metal layer partly reflects the probe pulse, which interferes with the reflection from the acoustic-wave-induced refractive index variation. The interference leads to the so-called Brillouin oscillations. Brillouin oscillations are one of the main features analyzed within the ps sonar signal, together with signal features connected to the acoustic wave arrival at an interface or surface (echoes).

Experiments commonly focus on the period of Brillouin oscillations to assess the index of refraction, timing of echoes for precise film thickness determination, amplitudes of echoes to detect delamination [3], or the shape of the echoes to detect interband transitions [8]. A part of the measured signal is also a long-lived decay induced by heat transfer in the thin films. However, the long-term decay is usually subtracted [7, 9].

While the signal in ps sonar has been extensively studied for many systems, so far, the signal has been used without considering the possible changes in the layers due to aging. Since it is well known that thin films in general undergo long-term aging comprising oxidation, stress relief, and water vapor adsorption [10-12], it is highly important to study this phenomenon in light of the ps sonar experiment.

In this article, we present a thorough study of the aging process in a thin-film stack studied by the ps sonar experiment. In particular, we studied the $Si_3N_4$ layer overcoated with a thin Ti layer serving as an acousto-optic transducer. We show that the measured signal changes on the scale of weeks and even months after the deposition. We study the changes in the signal with respect to the thermal-induced decay, timing of the echo features, and the period and phase of Brillouin oscillations.

We carried out a set of experiments to identify the origin of the changes, which tested the role of the ambient environment, laser irradiation, and diffusion of atoms across the interfaces. A part of our study also focused on the reversibility of these sample modifications and their suppression.

The article provides insight into the process of aging in the ps sonar experiment and the connected changes, which are important to consider, especially when focusing on subtle changes in the signal.

## 2. Methods

### 2.1. Ps sonar

Ps sonar setup was based on a fs amplified laser (100 kHz repetition rate, 225 fs pulses, 1028 nm). The infrared 1028 nm pulses served as the pump pulses (energy 1.5 µJ/pulse, spot size 400 µm), and probe pulses were attained via second-harmonic generation (514 nm, spot size 35 µm). The pump and probe beams were at a small angle (2 degrees), approaching the sample close to normal incidence. A balanced pair of photodiodes measured the differential signal between the reflected probe from the sample and the reference probe. Since the pump was modulated by an optical chopper, the differential signal analyzed via a lock-in amplifier provided the desired ps sonar signal. A simplified experimental setup is depicted in Fig. 1a.

Owing to the interference of probe beam reflections $r_1$ and $r_2$, shown in Fig. 1b, we observed the oscillations in the reflectivity signal depending on the current position of the acoustic wave, the so-called Brillouin oscillations from which the wave speed in the material and possibly the elastic modulus can be determined [13]. Echoes (returning of the part of the strain wave to the surface after reflection from the interface) and transitions across the layers' interfaces can be used to study the morphology of the interfaces. These features are labeled in the measured transient reflectance data in Fig 2.

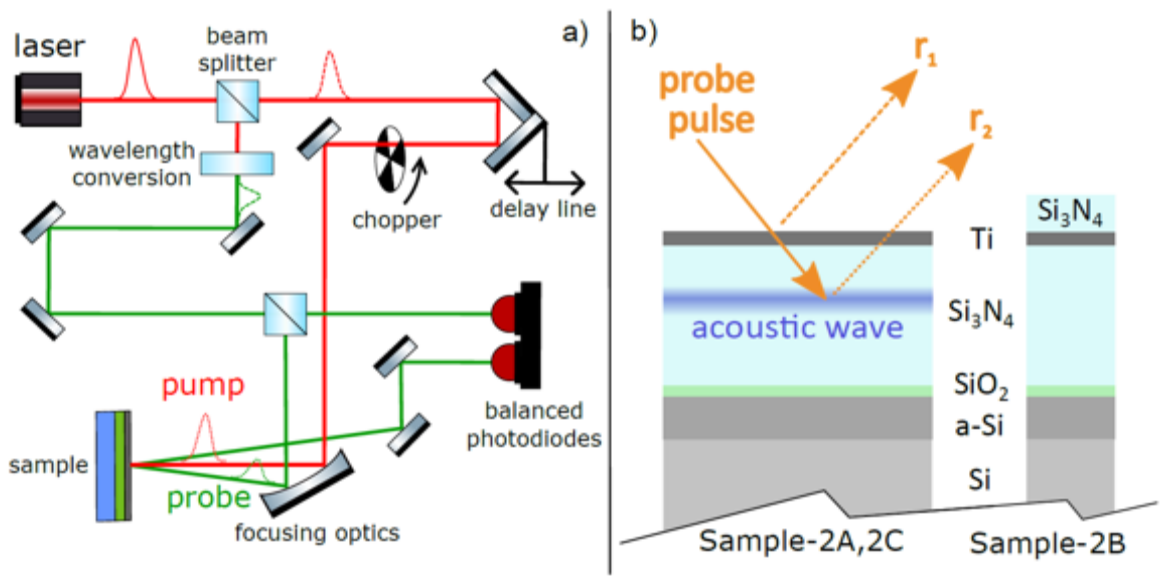

**Figure 1: Scheme of the experiment.** a) Experimental setup described in the text. b) The principle of the detection of the acoustic wave and the scheme of the studied samples; unpassivated *Sample-2A* and *2C* on the left and passivated *Sample-2B* on the right.

## 2.2. Thin film samples

The studied samples were prepared by dual-beam ion beam sputtering (IBS). Before depositions, the chamber was evacuated to a pressure smaller than $10^{-3}$ Pa and the substrate was cleaned using $Ar^+$ ions from an assistant ion gun. Ti and Si targets were sputtered onto the sample by a primary ion gun ($U_{beam}$ = 600 V, $I_{beam}$ = 108 mA). The sputtered Si ions either formed an amorphous Si layer (a-Si), or we used the assistant ion source to induce reaction with nitrogen ions to form $Si_3N_4$, or with the oxygen ions to form $SiO_2$. The assistance gun supplied concurrently also $Ar^+$ ions to add energy to the deposited ions. As a result, dense layers (index of refraction for the $Si_3N_4$ layer $n$ = 2.026 at the wavelength 500 nm) can be produced. See [14] for details of the deposition process.

We used p-type (1,0,0)-oriented crystalline Si (c-Si) wafers as a substrate. Directly on the substrate, we sputtered an amorphous Si (a-Si) layer. The goal of this layer was to eliminate strong Brillouin oscillation from the Si substrate hindering the echoes and interface features. Since sputtered a-Si highly absorbed the probe pulse, the Brillouin oscillation from the Si substrate are strongly attenuated, and the transitions of the wave across all the interfaces inside the multilayer could be clearly detected – see Fig. 2.

Since the samples were also deposited to study the effect of interlayers in the ps sonar signal, some of our samples intentionally include a thin $SiO_2$ interlayer deposited on the amorphous Si. The following layer was a relatively thick $Si_3N_4$ layer (hundreds of nanometers) followed by a thin (10-15 nm) Ti transducer layer. A scheme of the studied stacks is presented in Fig. 1b.

Throughout the text, we will denote the samples via the target thicknesses for the deposited layers. Since the layer thickness was monitored by a quartz crystal microbalance, the actual sputtered thicknesses can differ from the setpoints by ca. 2%. We checked the thickness of the relatively thick $Si_3N_4$ layer by ellipsometric measurements and confirmed the accuracy. For the Ti and $SiO_2$ layers the expected inaccuracy is below 1 nm. The thickness of amorphous Si was not an important parameter for the presented results.

In the initial part (Section 3.1), we present results from a sample denoted as *Sample-1*, which was designed as 10 nm Ti / 665 nm $Si_3N_4$ / 10 nm $SiO_2$ / 100 nm a-Si on c-Si substrate. The thickness of the $Si_3N_4$ layer determined by the ellipsometric measurements was 678 nm.

The major part of the results (Section 3.2) was carried out for samples prepared specifically to test the aging effect. For their preparation, we used the following procedure. Firstly, we deposited on a large Si substrate the basic stack (600 nm $Si_3N_4$ / 10 nm $SiO_2$ / 200 nm a-Si) without Ti layer. We denote this part as a principal stack.

One half of the large Si substrate with the principal stack was left aside in the ambient atmosphere. Subsequently, the other part was again divided into two halves; the first part, called *Sample-2A*, was covered with the bare metal Ti layer, as is usually used in ps acoustics. The second half, called *Sample-2B*, was covered with the Ti layer and, in addition, passivated by 10 nm of $Si_3N_4$ (see Fig. 1b).

Later, 192 days after the preparation of *Sample-2A* and *2B*, *Sample-2C* was prepared by using the same principal stack, which went through aging in ambient conditions. The aged principal stack was freshly covered by a new Ti layer with parameters identical to *Sample-2A*. The scheme of the samples is depicted in Fig. 1b.

At the end of the article, we show for comparison results from samples denoted as *Sample-3* and *Sample-4*, which are composed of 15 nm Ti / 611 nm $Si_3N_4$ on c-Si substrate and 10 nm Ti / 665 nm $Si_3N_4$ / 100 nm a-Si on c-Si substrate, respectively.
The composition of all the samples is summarized in Table 1.

**Table 1: Thicknesses of the individual layers in the samples presented in the article.** All samples were deposited on a Si substrate. The first column corresponds to the top layer exposed to air, and the last column is the layer adjacent to the substrate. Layers marked as "-" were not used for the particular sample.

| name | $Si_3N_4$ (nm) | Ti (nm) | $Si_3N_4$ (nm) | $SiO_2$ (nm) | a - Si (nm) |
|---|---|---|---|---|---|
| Sample-1 | - | 10 | 665 | 10 | 100 |
| Sample-2A | - | 15 | 600 | 10 | 200 |
| Sample-2B | 10 | 15 | 600 | 10 | 200 |
| Sample-2C | - | 15 | 600 | 10 | 200 |
| Sample-3 | - | 15 | 600 | - | - |
| Sample-4 | - | 10 | 665 | - | 100 |

# 3. Results

## 3.1. Changes in the transient reflectivity signal due to aging

During long-term studies of picosecond sonar samples, we commonly observed that the signal differed when measured after several weeks or even days. In Fig. 2 we show an example of a measurement of *Sample-1* using the same experimental parameters.

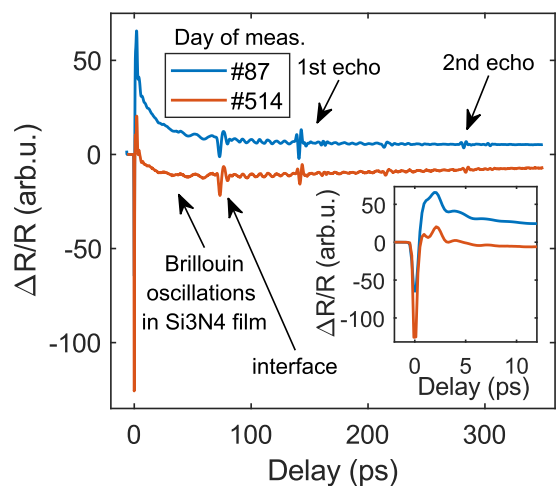

**Figure 2: Transient reflectivity signal of a thin multilayer *Sample-1* measured 3 months after the deposition and more than one year later.** The labels indicate important components of the ps sonar signal discussed in the text. Inset shows the detail of the first 12 ps of the dynamics.

Figure 2 shows apparent changes in several aspects of the measured signal. Firstly, the initial signal during the first 10 ps exhibited more pronounced modulation in the latter measurement. The initial signal is very complex, as it comprises an intermix of the optical response for the acoustic wave and the electron dynamics in the Ti layer itself. Therefore, we will focus in this article on the dynamics after 10 ps, where the signal interpretation in terms of the picosecond experiment is much more straightforward.

We observed that the shape of the long-term decay, connected to the thermal conduction in the sample, undergoes prominent changes. The exponential thermal background of the ps sonar signal is typically subtracted from the data [7, 9]. Therefore, these changes might not be visible in the commonly published data. However, the signal variation clearly indicates changes in the samples.

The presented kinetics were normalized with respect to the amplitude of the Brillouin oscillations in the $Si_3N_4$ layer between 41.4 ps and 62.2 ps. An analogous procedure was used throughout the presented results in this article to eliminate the influence of slight variation in laser fluence or pump-probe beam overlap. Since the initial dynamics varied highly with the sample aging and could not be used for the normalization, we used the amplitude of Brillouin oscillations as the normalization factor. Our measurement confirmed that the sample reflectivity at the probe wavelength 514 nm was changing over time by less than 1.5%. At the same time, we expect only negligible changes in the photoelastic properties of the $Si_3N_4$ layer, as we discuss later.

## 3.2. Controlled aging analysis

We prepared dedicated samples to study the process of aging changes starting from the sample's exposure to air. The sample was a common stack with a bare Ti layer on top (*Sample-2A*) and a sample with an additionally passivated Ti layer overcoated with 10 nm thick $Si_3N_4$ layer (*Sample-2B*). We measured the signal of a ps sonar on a regular basis from the first day of the preparation for 171 days in total.

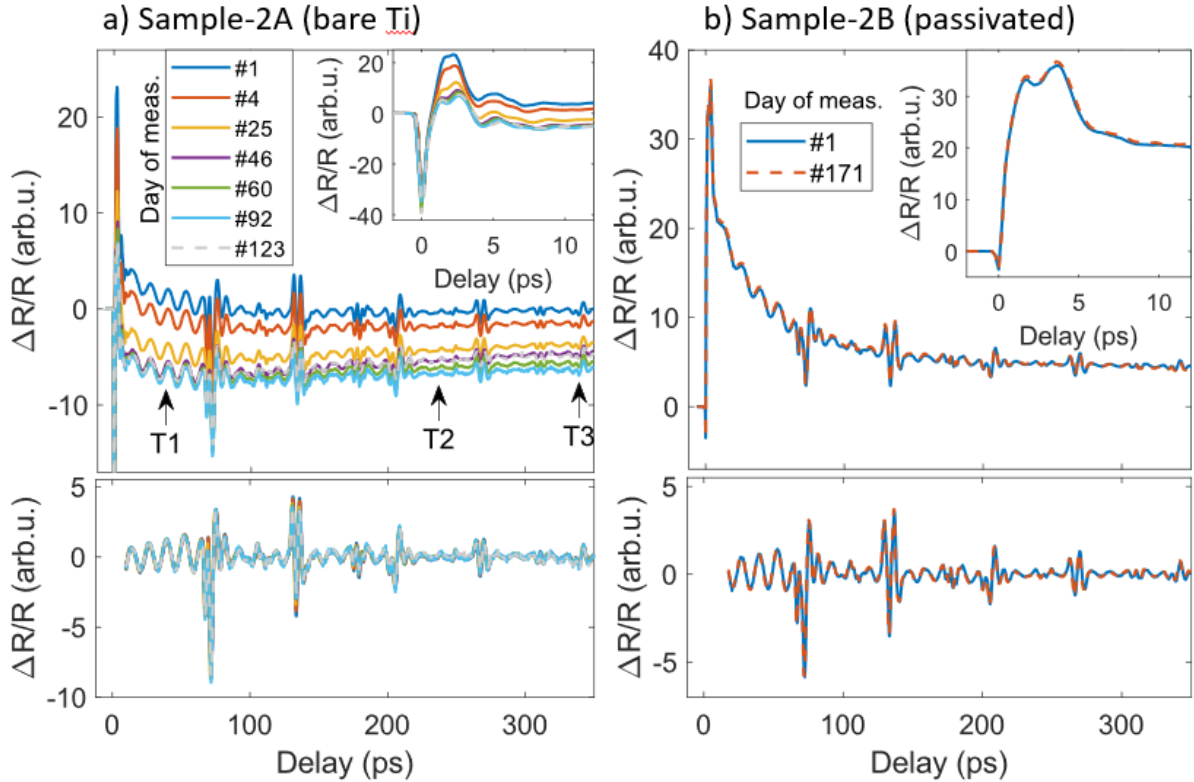

**Figure 3: Transient reflectivity signal of a) common unpassivated *Sample-2A* and b) passivated *Sample-2B* measured from the first day after deposition for 171 days in total.** The signal was normalized for the amplitude of Brillouin oscillations. For clarity reasons, only selected datasets are displayed. The data for day #123 corresponds to the state after staying in a vacuum, discussed in Section 3.2.1. The arrows in a) at time T1 and T2 indicate delays for the analyses in Fig. 4, the T3 arrow points to the position of the feature analyzed in Section 3.2.5. Insets show the detail of the initial dynamics. Bottom panels: the same data as in top panels after subtracting the double exponential fit of the long-term decay.

The traces of transient reflectivity measured for various ages of the samples, which were normalized with respect to their amplitude of the Brillouin oscillations in the $Si_3N_4$ layer, are shown in the top panels of Fig. 3.

In accordance with our previous measurements in Section 3.1. we observed that for the unpassivated *Sample-2A* (Fig. 3a) the long-lived decay (thermal background) systematically changed over time. Prominent changes appeared already a few days after the deposition. The ps sonar signal up to approx. 60 ps delay "stabilized" a month after the deposition. Nevertheless, for delays above 60 ps, the ongoing changes were apparent more than two months after the deposition. This indicates that we observe multiple mechanisms of the sample aging acting on a different timescale.

After subtracting a double-exponential fit from the measured dynamics (see bottom panels of Fig. 3a) we observe that the aging itself has only a subtle effect on the features connected to the acoustic echo and interface feature. We will address these differences later in Section 3.2.5.

Since we suspected the changes were connected with the Ti layer based on its exposure to air, a straightforward solution to the aging is to passivate the Ti layer itself by overcoating the layer with another thin Si$_3$N$_4$ film - *Sample-2B*. Indeed, we observed that the signal of the sample with the passivated Ti layer did not undergo any visible changes for nearly 6 months - see Fig. 3b.

The normalization enabled visualization of the relative changes in the dynamics by using a normalized ΔR signal at two selected representative delays: T1 = 43.6 ps and T2 = 238.6 ps, which are marked by arrows in Fig. 3a. The values of ΔR in these delays for both studied samples, *Sample-2A* and *Sample-2B,* are presented in Fig. 4. We will first focus on *Sample-2A* with an unpassivated Ti layer exposed to the ambient atmosphere. The descending values of ΔR in Fig. 4 can be fitted very well by an exponential function. The exponential fit for the values in the delay T1 has a characteristic time of 18.6 days, in the delay T2 of 25.3 days (orange and yellow dashed line in Fig. 4). On the other hand, the values of ΔR at delay 43.6 ps (point T1) for *Sample-2B* with a passivated Ti layer do not show any obvious dependence on the sample age. The dashed cyan line in Fig. 4a notes the mean of the values.

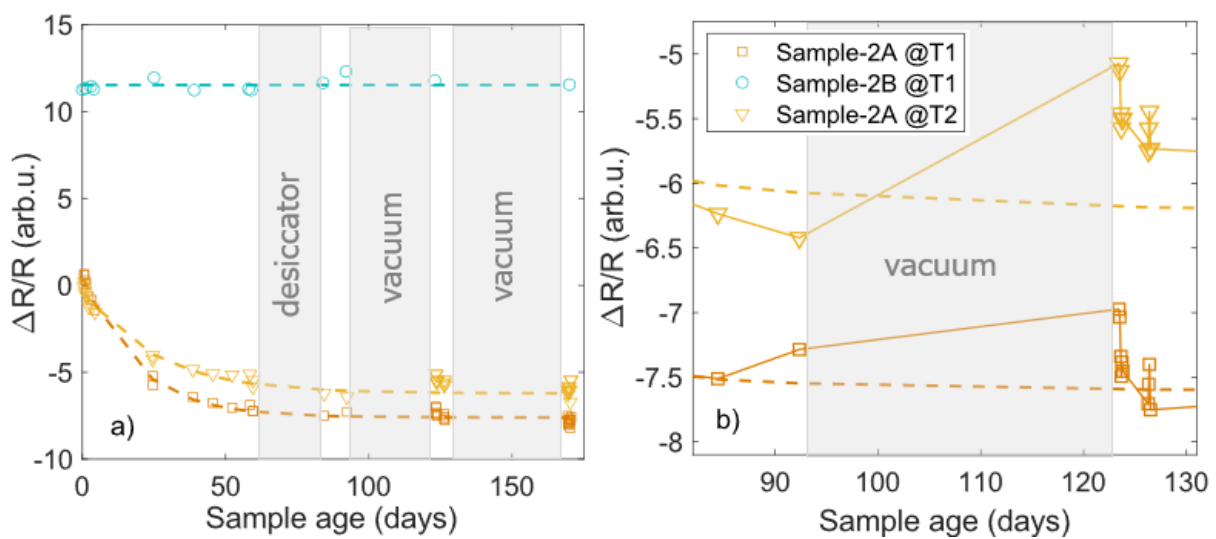

**Figure 4: Values of the relative reflection change measured in the delay time T1 = 43.6 ps and T2 = 238.6 ps after excitation (see Fig. 3a) as a function of the number of days after Ti layer deposition.** Grey rectangles indicate the periods when the sample was placed in the desiccator and in the vacuum. a) Whole measured scale. b) Detail of the changes after removal from vacuum. The legend is the same for both panels. Orange squares: unpassivated *Sample-2A* at delay T1, cyan circles: passivated *Sample-2B* at delay T1, yellow triangles: *Sample-2A* at delay T2. Cyan dashed line is the mean of all of the values for *Sample-2B*, orange and yellow dashed lines are exponential fits of the values before putting the *Sample-2A* into vacuum.

To study more closely the origin of the changes, we carried out several tests, where we placed the samples in different ambient conditions (Fig. 4a). Firstly, we placed the samples in a **desiccator**, where the samples stayed under low-humidity conditions in air for 25 days (day #59-84). This action had only a subtle effect on the signal of both samples – the value of ΔR in both studied delays for the unpassivated *Sample-2A* continued in exponential decrease, while the signal for the passivated *Sample-2B* did not change.

Secondly, the samples were placed for an extensive time of 31 days (day #92-123) and 43 days (day #126-169) in **vacuum**. For both vacuum tests, we observed for *Sample-2A* partial recovery of the aging, followed by abrupt nonmonotonous changes. The signal behavior after vacuum exposure was measured continuously over 3 days (see the enlarged detail of the dynamics after the first test in vacuum in Fig. 4b). Its complex shape confirms the presence of several mechanisms interplaying at the same time. Due to the effect of vacuum, we can identify water vapor and surrounding gas adsorption as viable sources of the observed changes.

Both samples, *2A* and *2B*, underwent the same environmental treating. Analogously to the aging itself, the environmental tests did not lead to any changes for the passivated *Sample-2B* (see cyan circles in Fig. 4a). The observed signal stability for the passivated sample implies that the observed changes cannot originate from the Ti atom diffusion into the $Si_3N_4$ layer. If this were the case, we would observe analogous signal changes for both passivated and unpassivated sample, since this diffusion would be present for both samples.

### 3.3. Verification measurements

To avoid possible artefacts or observation of locally specific defects, we carried out several verification measurements on *Sample-2A*. Firstly, we verified that the thin film is homogeneous and that the changes are not induced by the laser irradiation itself. The measurements were performed on several spots directly after deposition to prove that the sample was homogeneous (blue, red and yellow lines in Fig. 5) and again 171 days later. We measured on a "fresh" spot, which was not previously irradiated by the laser pulses (green dashed line in Fig. 5). Here, the ∆R dynamics showed the same shape, expected from the aging process. These two measurements were done after taking the sample out from vacuum, where the changes in time were very fast (see Fig. 4). Hence, also the one-hour difference in the measurement time was responsible for a small difference in the dynamics in the spots D and E.

Based on the results, we can exclude the possibility that the changes are laser-induced, and we confirmed that the sample undergoes homogeneous aging across its entire surface.

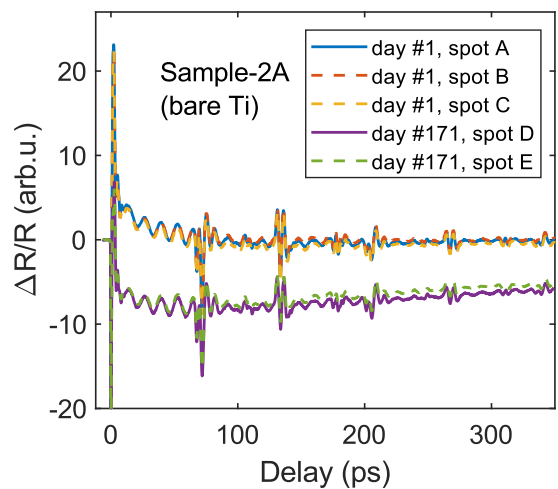

**Figure 5: Verification of homogeneity of the sample and aging process.** Blue, red and yellow lines: Transient reflectivity signal measured for *Sample-2A* with bare Ti layer directly after the deposition on three different spots. Purple and green lines: Transient reflectivity signal for the same sample measured 171 days after the deposition on two different spots.

## 3.4. Effect of fresh Ti layer

In order to pinpoint whether the aging originates from the Ti layer, which is exposed to the environment, or is connected to the $Si_3N_4$ layer below, we compare the transient reflectivity dynamics on a pristine $Si_3N_4$ layer (i.e. principal stack) overcoated with Ti (*Sample-2A*) with the same principal stack aged in air, which was later overcoated with a new Ti layer (*Sample-2C*).

As can be seen in Fig. 6, the shape of the signal of the aged multilayer with the fresh Ti overcoating (*Sample-2C*) is the same as that of *Sample-2A* directly after Ti deposition. Small differences between the curves can be explained by a difference in Ti layer thickness in *Sample-2A* and *2C*. We carried out the same tests, which were presented for *Sample-2A* in Fig. 3-4, also with *Sample-2C*. The changes of the dynamics over time follow the same trend as observed in S*ample-2A* (see Fig. 6, yellow line).

Based on the results, we can conclude that the changes depend on the age of the Ti layer and they are connected to this metal layer.

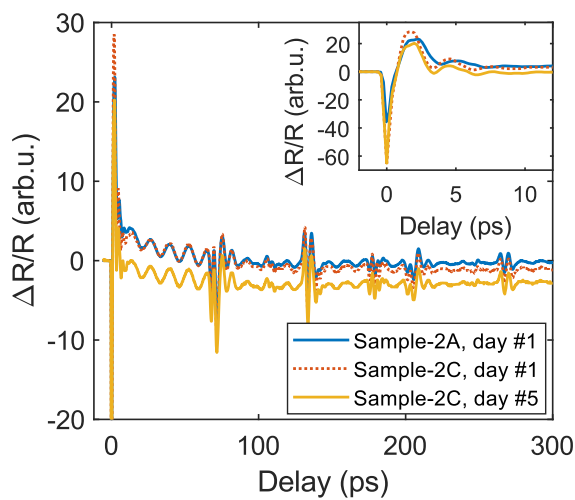

**Figure 6: Transient reflectivity signal measured on *Sample-2A*, i.e. Ti transducer layer deposited on a freshly prepared principal stack (blue line), and on *Sample-2C*, i.e. Ti transducer layer deposited nine months later on the other half of the same principal stack (red dashed line). Both signals are measured immediately after Ti deposition. Yellow line: *Sample-2C* measured 5 days after the Ti layer deposition. Inset: detail of the initial dynamics.**

## 3.5. Aging and the opto-acoustic features

In the text above, we focused on the thermal background in the ps-sonar signal, where the changes were the most apparent. Nevertheless, in light of ps sonar experiment, it is also highly important to evaluate the effect of the changes on the echoes and transitions through interfaces.

Owing to the sample design, where the strong optical response from the Si substrate was suppressed by a layer of amorphous Si, we were able to observe echoes also after multiple reflections in the stack. Hence, we were able to investigate the shape and the exact timing of the features in the ps sonar. Namely, we focused on the feature resulting from the acoustic wave passing through $Si_3N_4$ layer / $SiO_2$ interlayer interface after four preceding reflections from the interface and surface. In other words, the timing of this feature (340 ps; point T3 in Fig. 3) corresponds to five transfer times (68 ps) over the $Si_3N_4$ layer.

Figure 7a and 7b show the detail of the ps sonar signal in delays 300-350 ps, including the above-mentioned transition of the acoustic wave through the layer/interlayer interface around the delay of 340 ps and also the echo of the acoustic wave traveling four times in the $Si_3N_4$ layer and two times in the a-Si layer (i.e., it is reflected from both interfaces - layer/interlayer and a-Si/c-Si), which arises around 310 ps. As the $SiO_2$ interlayer is so thin, the transfers of the acoustic wave through both interfaces of this interlayer are merged in one feature. We plot the data measured the first day after Ti deposition and the last, 171$^{st}$, day of the measurement for the unpassivated *Sample-2A* and for the passivated *Sample-2B*, respectively.

The signal features (days #1 and #171) for the unpassivated *Sample-2A* smeared out, while, on the contrary, for the passivated *Sample-2B*, both curves had identical shapes. We only observed that the delay between the two features differed by 200 fs. However, the subtle difference of 0.2 ps at the delay of 340 ps can be well explained by a small change in the local temperature.

Since it might be problematic to analyze the exact feature timing for a varying shape, we searched for the position of the maximum of cross-correlation between the feature trace on the first scan (day #1, reference trace) and the following scans. For each scan, we determined the exact timing difference of the feature - see Fig. 7c and d. The timing varied in the range of ±400 fs without any reliable systematic trend. Nevertheless, based on the fit, we can state that the feature timing due to the layer aging varied by less than 0.3 ps. Since we studied the signal at the delay of 340 ps, we can conclude that the speed of longitudinal acoustic waves changed during the aging by less than 0.1%.

The same conclusions can be drawn also from the period of Brillouin oscillations in the initial ps sonar dynamics (delays up to 60 ps) – see Fig. 8. We fitted the data with a sine function superimposed on a linear background, which was important to account for the remaining long-term drift. Here the period of oscillations reached (12870±30) fs for Sample-2A, and (12830±30) fs for Sample-2B. The period remained constant throughout the measurements. Since the period is given both by the speed of sound and the layer refractive index, we conclude that also the variation in the layer refractive index was below 0.2%. This is in a very good agreement with the measurement of reflectance, where the shift in the interference pattern caused by the stack aging corresponded to the shift in the total optical path, i.e. refractive index, by 0.15%.

The results concerning feature timing and Brillouin oscillation period are again consistent with the aging taking place dominantly in the thin surface Ti layer. The changes in the Ti layer have no effect on the Brillouin oscillation period, as it was observed for our samples, and only small effect on the timing of ps sonar features, which might be easily hindered by other effects - for instance, changes in local sample temperature due to the laser beam spot changes.

At the same time, we observed a gradual slight change in the phase of the Brillouin oscillations for the unpassivated *Sample-2A*. Such behavior can be explained by a change in the refractive index of a thin Ti layer on the sample surface, which will not affect the Brillouin oscillations period – only the phase due to the change in surface reflection contributing to the interference.

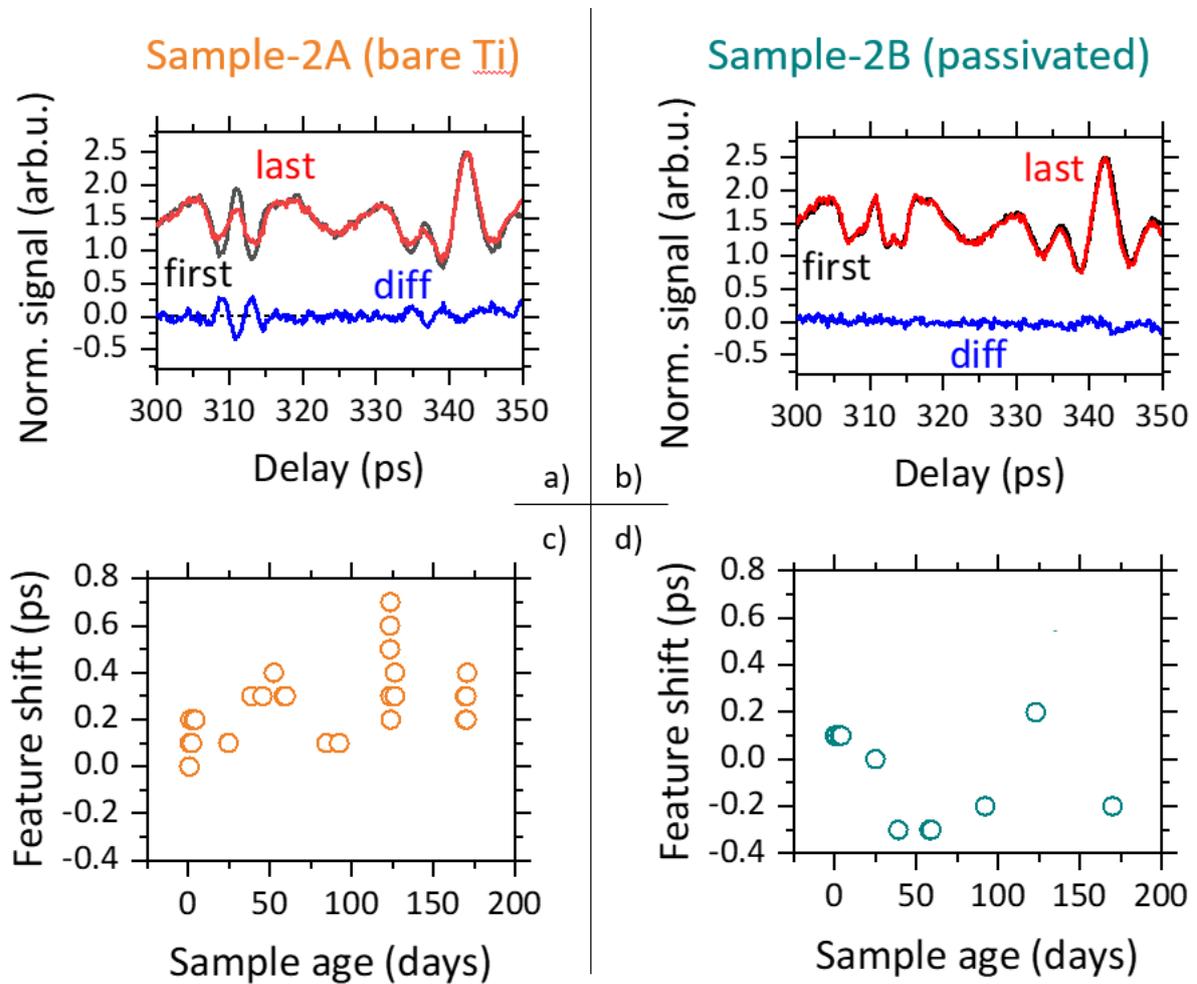

**Figure 7: Transient reflection signal for the point T3** (see Fig. 3) for day #1 (first – black line) and for day #171 (last – red line) with their difference (diff – blue line). Unpassivated sample on the left, passivated on the right. Note that an offset was added to the signal to avoid overlap with the "diff" line. a)-b) The shape of the signal, c)-d) Shift in feature delay compared to day #1 signal, given as a maximum of the cross-correlation between the individual measurements.

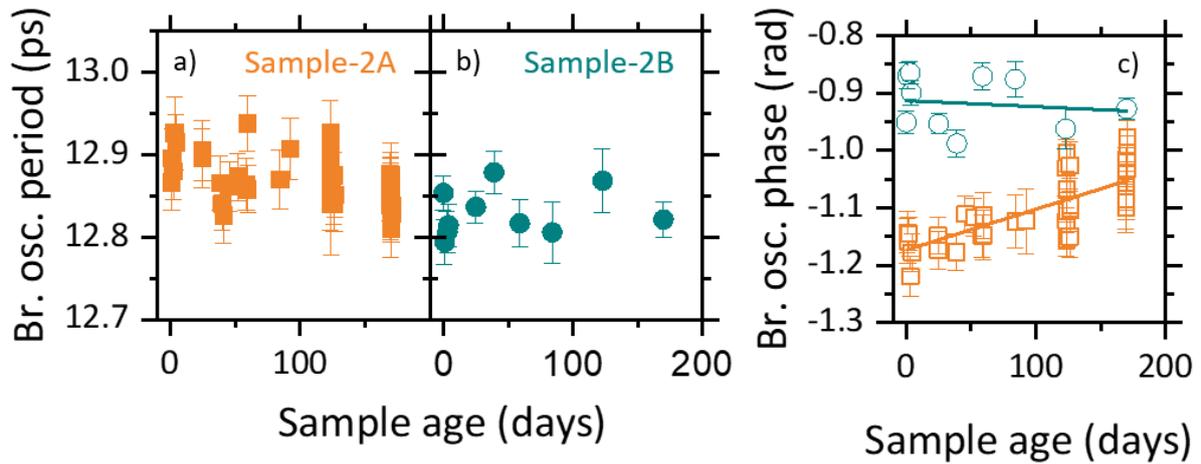

**Figure 8: a)-b) Period and c) phase of the Brillouin oscillations** obtained from signal fitting for delays 22 – 50 ps by a sine function with a linear background. Orange squares stand for unpassivated *Sample-2A*, green circles for passivated *Sample-2B*. The error bars are the std errors of the fit.

## 4. Discussion

Our measurements and analyses show that the changes we observed in the ps sonar signal do not take place directly in the $Si_3N_4$ layer. Firstly, we verified that the changes are connected to the Ti layer's exposure to air and humidity and can be avoided by passivating the Ti layer. Secondly, the speed of the acoustic wave and the index of refraction in $Si_3N_4$ layer did not noticeably change, while the phase of the oscillation underwent a slight shift. At the same time, the observed changes cannot originate from the Ti atom diffusion into the $Si_3N_4$ layer, since this diffusion would be present for both passivated and unpassivated samples. Therefore, the main suspect is the Ti layer itself.

A commonly reported source of aging is the penetration of atmospheric water into the thin films. In this case the aging is expected to be reversible in a vacuum, as we observed. According to Ref. 11, water adsorption was observed to change the optical thickness of the $HfO_2/SiO_2$ coatings by ~2.2% for humidity change from 1% to 55% or, in more realistic case for us, by ~0.5% for humidity change from 30% to 55%. In the case where water would be adsorbed also in the $Si_3N_4$ layer, we would expect the frequency of Brillouin oscillations to shift accordingly. Based on the fact that no such shift was observed, we can conclude that significant water adsorption can be present only in the Ti layer or a minimal portion of the $Si_3N_4$ layer, which is not sufficient to vary the total optical path in the layer. Moreover, based on the fact that 10-nm $Si_3N_4$ layer effectively blocked Ti layer aging, we can expect that such water adsorption would be restricted only to several nanometers of the $Si_3N_4$ layer itself.

Another viable source of Ti layer changes is its oxidation. The Ti oxidation is usually very rapid. A few-nm thick $TiO_2$ layers [15] are created within a few hours, even at room temperature and even for later slower processes, bound oxygen content was observed to saturate within 4 days (100 hours) [12]. Such effect can account for the changes within the first days after deposition, but not on the longer timescale.

Nevertheless, according to Logothetidisa et al. [12], even though later oxygen does not react with the film, it still diffuses into the intercolumnar structure of the film and changes the stress in the layer. This process was reported to not saturate even after 500 hours (20 days). Such behavior is in line with our experiment, when we placed the sample in a desiccator. There, the oxygen was still present and, as a result, despite the low humidity, the aging process was not stopped. A similar effect of long-term stress variations in thin films was also observed in our experiments via the second-harmonic generation technique [16].

Final strong evidence that aging affects the Ti layer itself are the prominent changes in the initial signal less than 5 ps after excitation. In this timeframe, the acoustic wave is highly affected by the surface Ti layer itself. Moreover, the transient reflectivity signal is influenced also by relaxation of electrons in Ti.

In Fig. 9, we show the detail of the initial signal and how it evolved in the studied sample. Depending on the sample, we observed that aging led to pronouncing the initial oscillations, while some samples evolved in the opposite way. As the initial signal is very complex, it consists of a combination of the optical response for the acoustic wave and the electron dynamics in the Ti layer itself, we were not able to interpret the changes. An exact theoretical numerical model, which is currently being developed, might shed light on the issue in the future.

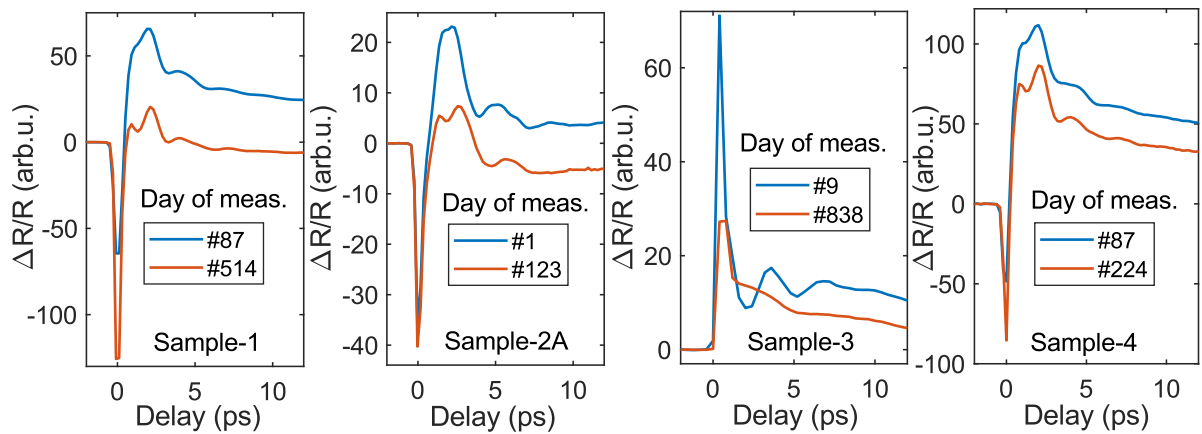

**Figure 9: Detail of first few ps of the signal for all presented unpassivated samples.**

## 5. Conclusion

To conclude, we observed prominent changes in the measured transient reflectivity signal of picosecond sonar, which rapidly evolved during the first days after the sample deposition and the changes continued over the scale of weeks and months. By designing a set of experiments with passivated and unpassivated samples, as well as using a "new" Ti layer on "old" thin film stack, we proved that the changes are connected to the Ti layer, which is commonly used as an opto-acoustic transducer in ps sonar experiment. Our results excluded the option that Ti atom diffusion to a layer below was responsible for the aging. The Ti layer exposure to the ambient atmosphere was identified as the key factor.

The changes were partially reversible in vacuum, but not in the desiccator - suggesting a combination of two or more processes. We identified the Ti oxidation (fast, non-reversible) and gas/water vapor adsorption (reversible in vacuum) as the most viable sources of the changes. The most likely effect responsible for the long-term aging on the timescale of weeks is the variation of the internal stress due to the water or oxygen diffusion into the pores. The very long time without observable saturation (171 days) and negligible effect of stay in the desiccator supports oxygen adsorption as a factor. Nevertheless, as only part of the changes were reversible, it is likely that more effects are interconnected in the observed process.

We verified that the changes in the sample were homogeneous across the sample surface and were not induced by the laser irradiation itself – the same aging effect was observed on the fresh spot, where the laser had not been incident.

The main studied features in ps sonar – Brillouin oscillations and echoes – were only partly affected by aging. Changes in the velocity of sound in the sample were proven to be below 0.1%. While the frequency of the Brillouin oscillations in the $Si_3N_4$ film did not change, their phase did, which again points to the origin of the effect in the Ti layer. The change in timing of the acoustic echoes and other features was negligible compared to other effects. Nevertheless, the shape of the features varied with sample age.

Therefore, the described effect is important to consider for studies of samples in ps acoustics, especially when focusing on subtle features in the signal. We stress that the most prominent changes occur already during the first days after deposition.

The key message of this work is that all the changes can be suppressed by overcoating the Ti layer by a thin $Si_3N_4$ layer. Using just a 10-nm $Si_3N_4$ layer, the effect of aging was canceled and samples were stable.

The data that support the findings of this study are available from the corresponding author upon reasonable request.

# Acknowledgement

This work was supported by Grantová Agentura České Republiky (23-08020S).

# Author contributions

All authors contributed to the manuscript equally.

# Competing interests

The authors declare no competing interests.

# References:

[1] H. Angus Mecleod, "Thin-film Optical Filters", Fourth Edition (CRC Press, Taylor & Francis Group, LLC, 2010).

[2] O. Matsuda, M. C. Larciprete, R. Li Voti, O. B. Wright, "Fundamentals of picosecond laser ultrasonics." Ultrasonics **56**, 3-20 (2015)

[3] A. Devos, A. Vital-Juarez, A. Chargui, M.J. Cordill, "Thin-film adhesion: A comparative study between colored picosecond acoustics and spontaneous buckles analysis," Surface and Coatings Technology **421**, 127485 (2021)

[4] Tas, G; Maris, HJ, "Electron-diffusion in metals studied by picosecond ultrasonicss" Phys. Rev. B **49** (21) 15046-15054 (1994)

[5] Akimov, AV; Scherbakov, AV; Yakovlev, DR ; Bayer, M" Picosecond acoustics in semiconductor optoelectronic nanostructures" Ultrasonics **56**, 122-128 (2015)

[6] Audoin, B; Rossignol, C ; Chigarev, N; Ducousso, M ; Forget, G; Guillemot, F; Durrieu, MC "Picosecond acoustics in vegetal cells: Non-invasive in vitro measurements at a sub-cell scale" Ultrasonics **50** (2) 202-207 (2010)

[7] O. Matsuda, O. B. Wright, "Reflection and transmission of light in multilayers perturbed by picosecond strain pulse propagation," J. Opt. Soc. Am. B **19** (12) 3028 (2002).

[8] A. Devos, R. Côte, "Strong oscillations detected by picosecond ultrasonics in silicon: Evidence for an electronic-structure effect," Phys. Rev. B **70**, 125208 (2004)

[9] C. Thomsen, H. T. Grahn, H. J. Maris, J. Tauc, "Surface generation and detection of phonons by picosecond light pulses," Physical Review B **34** (6) 4129 - 4138 (1986)

[10] Parada, E. G. et al. "Aging of photochemical vapor deposited silicon oxide thin films." J. Vac. Sci. Technol. A **14**, 436–440 (1996)

[11] J.F. Anzellotti, D.J. Smith, R.J. Sczupak, Z.R. Chrzan, "Stress and environmental shift characteristics of HfO2/SiO 2 multilayer coatings," Proceedings of SPIE - The International Society for Optical Engineering **2966**, 258 – 264 (1997)

[12] S. Logothetidisa, E.I. Meletisb, G. Stergioudisa, A.A. Adjaottorb, "Room temperature oxidation behavior of TiN thin films" Thin Solid Films **338**, 304-313 (1999)

[13] A. Devos, "Colored ultrafast acoustics: From fundamentals to applications," Ultrasonics **56**, 90–97 (2015)

[14] V. Kanclir, J. Václavík, and K. Zídek, "Precision of Silicon Oxynitride Refractive-Index Profile Retrieval Using Optical Characterization," Acta Phys. Pol. A **140**, 215–221 (2021).

[15] V A Matveev, N K Pleshanov, A P Bulkin, V G Syromyatnikov, "The study of the oxidation of thin Ti films by neutron reflectometry." Journal of Physics: Conference Series **340**, 012086 (2012)

[16] Lukeš, J., Kanclíř, V., Václavík, J., Melich, R., Fuchs, U., Žídek, K., "Optically modified second harmonic generation in silicon oxynitride thin films via local layer heating." Scientific Reports **13**(1), 8658 (2023).